%% file: sp01cp.tex
\documentstyle[11pt]{article}
\begin{document}
\title{\hfill
\parbox[l]{4.5cm}{\normalsize ANL-HEP-CP-02-002 \\
\normalsize\today}\\
\vspace{1cm}
POLARIZED PARTON DISTRIBUTIONS AND\\
THE POLARIZED GLUON ASYMMETRY}
\author{GORDON P. RAMSEY\thanks{E-mail address: gpr@hep.anl.gov. 
Talk given at SPIN 2001, Beijing, China.} \\
Physics Department, Loyola University Chicago, 6525 N. Sheridan\\
Chicago, IL 60626, USA \\ \\
and \\ \\
High Energy Physics Division, Argonne National Laboratory\\
Argonne, IL 60439, USA}

\maketitle

\begin{abstract}
The flavor-dependent valence, sea quark and antiquark spin distributions can 
be determined separately from theoretical assumptions and experimental data. 
We have determined the valence distributions using the Bjorken sum rule and
have extracted polarized sea distributions, assuming that the quarks and 
anti-quarks for each flavor are symmetric. Other experiments have been proposed
which will allow us to completely break the SU(3) symmetry of the sea flavors.
To create a physical model for the polarized gluons, we investigate the gluon
spin asymmetry in a proton, $A_G(x,Q^2)={{\Delta G(x,Q^2)}\over {G(x,Q^2)}}$.
By assuming that htis is is approximately $Q^2$ invariant, we can completely 
determine the $x$-dependence of this asymmetry, which satisfies
constituent counting rules and reproduces the basic results of the 
Bremsstrahlung model originated by Close and Sivers. This asymmetry can be
combined with the measured unpolarized gluon density, $G(x,Q^2)$ to provide a
prediction for $\Delta G(x,Q^2)$. Existing and proposed experiments can test
both the prediction of scale-invariance for $A_G(x,Q^2)$ and the nature of
$\Delta G$ itself. These models will be discussed along with suggestions for 
specific experiments which can be performed at energies typical of HERA, RHIC 
and LHC to determine these polarized distributions.

\end{abstract}

\section{Introduction}

One of the important questions in high energy physics is how the quark and
gluon constituents contribute to nucleon spin. Significant interest in high
energy polarization was generated when the European Muon Collaboration
(EMC)\cite{emc} analyzed polarized deep-inelastic scattering (DIS) data which
implied that the Bjorken sum rule (BSR) of QCD\cite{bj} was satisfied and the
Ellis-Jaffe sum rule\cite{ej} based on a simple quark model was violated.
Since then, the Spin Muon Collaboration (SMC)\cite{smc}, experimental groups
from SLAC\cite{slac} and the HERMES group\cite{hermes} have measured $g_1$ to
lower $x$ with improved statistics and have lowered the systematic errors from
the original data.

Our general approach has been to split the polarized quark distributions into
valence and sea components, and to use theoretical constraints and data to
determine these distributions.\cite{ggr} Theoretical constraints include the 
Bjorken sum rule, the quark counting rules at large $x$ and positivity at 
leading order (LO). The structure functions $g_1$ for the proton, neutron and 
deuteron can be extracted from the corresponding polarized DIS asymmetries 
$A_1$, by assuming that the structure functions $g_2$ are small and that $A_1$
is relatively independent of $Q^2$. Additional experimental information is 
extracted from hyperon decay data. The $Q^2$ dependence of the distributions 
is generated by next-to-leading order (NLO) evolution.

The purpose of this paper is to outline present polarized quark models, 
introduce a gluon model and discuss how these can be determined with 
experiments planned at the major accelerators. The talk is outlined as follows:
(1) models of the polarized valence and sea quark distributions are summarized,
(2) means by which we can extract information about the polarized quark 
distributions from experiments are discussed (3) a physical model for the
polarized gluons via the asymmetry $\Delta G/G$ is presented and (4) a set of 
experiments is suggested, which would distinguish the quark and gluon 
contributions to the proton spin. 

\section{Quark Distributions}

\subsection{Valence quarks}

There have been two models proposed for construction of the valence quark
distributions. The original Carlitz-Kaur\cite{ck} model, based upon a modified
SU(6) quark configuration and the Isgur\cite{isgur} model, constructed from 
hyperfine splitting of the constituent quark model. 

The original Carlitz-Kaur model constructed the polarized valence quark 
distributions from the unpolarized ones by starting with a modified 3-quark 
model based on an SU(6) proton wave function. From this, the valence quark 
distributions can be written as:
$$\Delta u_v (x,Q^2)=\cos \theta_D [u_v(x,Q^2)-{2\over 3}d_v(x,Q^2)],$$
\begin{equation}
\Delta d_v (x,Q^2)=-{1\over 3}\cos \theta_D d_v(x,Q^2), \label{vq}
\end{equation}
where $\cos \theta_D$ is a "spin dilution" factor which vanishes as $x\to 0$
and becomes unity as $x\to 1$, characterizing the valence quark helicity
contribution to the proton.\cite{ck} Since the spin dilution factor is not 
derived from first principles, it is adjusted to satisfy the Bjorken sum rule,
which is considered to be a fundamental test of QCD. 
This enables us to determine the valence distributions explicitly.

The resulting valence distributions are not very sensitive to the unpolarized 
distributions used to generate them.\cite{ggr} In this model, 
$\langle \Delta u_v\rangle=0.90\pm 0.03$ and $\langle \Delta d_v\rangle=-.25
\pm 0.03$, the spin contribution from the valence quarks is $0.65\pm 0.06$. 
The errors arise from data on $g_A/g_V$ and higher order corrections. These
results are consistent with the measured magnetic moment ratio, $\mu_p/\mu_n$.

The Isgur model uses the hyperfine interactions of the constituent quark model
to predict the valence distribution in the kinematic region $0.3\le x \le 0.9$,
where valence quarks dominate in the proton. The shape of the polarized 
valence distributions are slightly different from the Carlitz-Kaur model, but
the essential features are the same. The minor differences are likely not 
distinguishable by any proposed experiments. Both models are consistent with 
recent SMC\cite{smc2} and HERMES data.\cite{hermes}

\subsection{Sea quarks}

We assume that the lightest flavors should dominate the spin of the sea, since
the heavier quarks would be significantly harder to polarize. The SU(6) 
symmetry of the sea can be completely broken by considering the following:

\begin{itemize}
\item assuming that the polarization of
the heavier strange quarks is suppressed,\cite{ggr} 
\item since the unpolarized quark flavors are asymmetric and we assume that the 
polarized distributions depend upon these, there is reason to believe that 
there is flavor asymmetry in the polarized quarks,\cite{kumano,morii}
\item chiral quark model predictions.\cite{wakamatsu}
\end{itemize}

The sea distributions are then related by: 
\begin{eqnarray}
\Delta \bar{u}(x,Q^2)&=&c_1\Delta u(x,Q^2)=c_2\Delta \bar{d}(x,Q^2)=c_3
\Delta d(x,Q^2) \label{sea} \\
&=&[1+\epsilon]\Delta \bar{s}(x,Q^2)=[1+\epsilon]\Delta s(x,Q^2),
\end{eqnarray}                                         
where $\epsilon$ is a measure of the increased difficulty in polarizing
the strange quarks and the $c_i$ are due to the asymmetries in the quark and
antiquark polarized distributions for each flavor. It is implicitly assumed 
that the strange sea is equally polarized ($\Delta s\approx \Delta \bar{s}$).
Since both are likely small, any differences due to meson-baryon effects in
the fluctuation of the proton wave function,\cite{mb} are likely not
distinguishable by polarized experiments. We also assume the the charm 
contribution to proton spin in this kinematic region is negligible compared 
to the light quark contributions.

Additional constraints are provided by the axial-vector current
operators, $A_3$, $A_8$ and $A_0$.
The coefficient $A_8$, determined by hyperon decay, 
$A_0$ is related to the total spin carried by the quarks
Through these relations, we can extract specific information about
individual contributions to the overall proton spin.

\subsection{Extraction of quark distributions from data}

There are various experiments proposed or in progress to extract the polarized
quark distributions. The valence models can be tested to a rough approximation
in deep-inelastic scattering\cite{hermes} if certain simplifying assumptions 
are made about the symmetry of the polarized sea. If we wish to extract the
separate flavors, independent of these assumptions, the best candidates would be
differences in asymmetries measured in $\pi^\pm$ and $K^\pm$ 
production.\cite{compass} The sea distributions and dependence of the 
asymmetries on the fragmentation functions cancel in these differences, so 
that the individual valence distributions may be found. Here
\begin{eqnarray}
A_p^{\pi^+}-A_p^{\pi^-} &=& {{4\Delta u_v-\Delta d_v}\over{4u_v-d_v}} \\
A_p^{K^+}-A_p^{K^-} &=& {{\Delta u_v}\over{u_v}} \\ \label{ap}
A_d^{\pi^+}-A_d^{\pi^-} &=& {{\Delta u_v+\Delta d_v}\over{u_v+d_v}},
\end{eqnarray}
where $p$ refers to a proton target and $d$ to a deuteron target. The proton
asymmetries would be sufficient to uniquely determine the valence 
distributions, but the deuteron measurement provides a good consistency check
for these models.

As shown in ref. [7], we can extract values of the spin contribution 
of the sea quarks from deep-inelastic scattering and hyperon data, provided we
assume symmetry within each flavor of the sea. Refer to Table 1 for a summary
of these results. The estimated errors are based on which model of $\Delta G$
is used and on the variations of the integrated structure functions 
($\int_0^1 g_1\>dx$) taken from each experiment.

\footnotesize
\begin{center}
{Table 1. Integrated Polarized Distributions} \\
\end{center}
$$\begin{array}{ccccc}
\hline
  Quantity         &Range        &Average &\Delta G\>unc. &Exp.\>Var. \cr
                   &             &        &              &         \cr
\hline
  <\Delta u>_{tot} &0.80\to 0.90 &0.86    &\pm 0.02      &\pm 0.04 \cr
  <\Delta d>_{tot} &-.36\to -.45 &-.40    &\pm 0.02      &\pm 0.04 \cr
  <\Delta s>_{tot} &-.02\to -.12 &-.06    &\pm 0.02      &\pm 0.04 \cr
  <\Delta q>_{tot} &0.23\to 0.52 &0.40    &\pm 0.02      &\pm 0.04 \cr
\hline
\end{array}$$ 

\normalsize
\vskip3mm

From Table 1, we can draw the following conclusions: 

(1) The naive quark model is not sufficient to explain the proton's spin
characteristics, since the total quark contribution to proton spin falls
between about $1\over 4$ and $1\over 2$, with the average about $1\over 3$.

(2) All flavors of the sea are clearly polarized. The up and down contributions
agree to within a few percent. All flavors satisfy the positivity bound
\cite{pos}. However, the widest percentage variation is found in the 
polarized strange sea.

(3) Although the flavor contributions to the proton spin cannot
be extracted precisely, the range of possibilities has been substantially
decreased. The main differences are the questions of the strange sea spin
content and the size of the polarized gluon distribution.

If we are to allow a a complete symmetry breaking of the sea, however, either 
more restrictions or different measurements are required for determination of
the different distributions. One set of possibilities comes from measuring the
$g_1$ and $g_5$ structure functions in charged current events, particularly 
$W^\pm$ production.\cite{ans} Polarized lepton pair production could also 
yield useful information on the spin contributions from the sea.\cite{morii}

One scenario would be to extract $g_1^+$ and $g_5^+$ from $W^+$ production and
$g_1^-$ and $g_5^-$ from $W^-$ production, in addition to the ratio of the 
integrated distributions, $\Delta u_{tot}/\Delta d_{tot}$ in lepton pair 
production. This provides five constraints to determine the five polarized
distributions: $\Delta u_s$, $\Delta d_s$, $\Delta \bar u$, $\Delta \bar d$ and
$\Delta s=\Delta \bar s$. The only significant unknown remaining is the size
and shape of the polarized gluon distribution.

\section{The Polarized Gluon Distribution}

The spin-weighted gluon density, $\Delta G(x,Q^2)$, is of fundamental
importance in understanding the dynamics of hadron structure. Numerous
experiments have been proposed\cite{hermes,compass,rhic} to extract this 
distribution. Although the polarized quark distributions are somewhat well 
known, the shape and size of $\Delta G(x,Q^2)$ has not been determined. There
have been phenomenological models proposed,\cite{ggr} but they are not based
upon strict physical models. However, the constituent quark model provides a 
framework for predicting an essential feature of $\Delta G(x,Q^2)$. 

Spin observables at small $Q^2$ conform to the non-relativistic quark model in
which spin degrees of freedom are associated with constituent quarks. Thus the
proton does not have a valence or ``constituent" gluon polarization. At low 
$Q^2\le m_P^2$, a proton consists of three ``valence" quarks, surrounded by 
radiated gluons and $q\bar q$ pairs. In a variation of the Close-Sivers 
Bremsstrahlung model\cite{cs}, gluons obtain their polarization from the 
valence quarks at low to medium values of $x$ and $Q^2$. The QCD evolution 
equations can then be used to generate a prediction for the quark and gluon 
distributions at higher $Q^2$, from a $Q_0^2$ where the constituent quark 
picture is applicable.

Using the assumptions from this model, we examine the gluon polarization 
asymmetry, defined as
\begin{equation} 
A_G(x,t)\equiv \Delta G(x,t)/G(x,t), \label{ag}
\end{equation}
where the evolution variable $t\equiv \ln[\alpha_s(Q_0^2)/\alpha_s(Q^2)]$.
It is assumed that the same factorization prescription is used to define all
of the densities in Eq. (\ref{ag}). 

In the absence of a ``constituent" gluon, both $G(x,t)$ and $\Delta G(x,t)$ 
exhibit scaling violations which can be associated with radiative diagrams. 
Since the positive and negative helicity gluon diagrams are the same. the 
probability of measuring a gluon of either helicity does not depend upon $t$.
Thus, we assume that the gluon polarization asymmetry is scale invariant: 
$\partial A_G(x,t)/\partial t=0$, where $t=0$ coincides with a typical 
hadronic scale, $Q^2=m_H^2$. Scale-invariance provides the $x$ dependence of 
$A_G(x)$, satisfying several physical constraints:
\begin{itemize}
\item it obeys the constituent-counting rules,
\item for large $x$, where quark contributions dominate the gluons, the 
asymmetry coincides with the original QCD-Bremsstrahlung model of Close and 
Sivers\cite{cs}. At other values of $x$, it corresponds to a natural extension
of this approach by allowing for radiation from both quarks and gluons, and
\item for small $x$, where the gluon contribution is dominant, the 
scale-invariant asymmetry has a natural asymptotic limit, independent of the 
starting point. 
\end{itemize}

The asymmetry, combined with parametrizations of polarized and unpolarized 
distributions provides an estimate for $\Delta G(x,Q_0^2)$ at any convenient 
reference scale. The requirement that $A_G(x,t)$ has no $t$-dependence implies
that
\begin{equation}
{{\partial A_G}\over {\partial t}}={1\over G}\Bigl[{{\partial \Delta G}
\over {\partial t}}-A_G(x,t) {{\partial G}\over {\partial t}}\Bigr]=0. 
\label{gpr:2.1}
\end{equation}
The $t$-dependence of the gluon distributions is given by the corresponding
DGLAP evolution equations. Since $\Delta G$ has not been measured, we can 
convert Eq. (\ref{gpr:2.1}) into a non-linear equation for $A_G(x)$ by 
inserting $\Delta G(x,t)=A_G(x)\cdot G(x,t)$ into the convolution of the 
evolution equations, giving
\begin{equation}
A_G={{{\partial \Delta G}\over {\partial t}}\over {{\partial G}\over
{\partial t}}}=\Biggl[{{\Delta P_{Gq} \otimes \Delta q+\Delta P_{GG}\otimes
(A_G\cdot G)}\over {P_{Gq} \otimes q+P_{GG}\otimes G}}\Biggr]. \label{gpr:2.2}
\end{equation}

An equation in this form can be solved iteratively. We first observe that for
a given value of $x$, the distributions in the DGLAP equations enter only in
the range $[x,1]$. Then, for a large enough $x$ ($x\geq 0.6$), the gluons can 
be neglected. The polarized DIS data are consistent with the constituent 
counting rule result that $\lim_{x\to 1}\> A_1(x,Q^2)\approx \lim_{x\to 1}
\>\Delta u_v(x,Q^2)/u_v(x,Q^2)=1.$ Thus, we make an initial approximation
\begin{equation}
\lim_{x\to 1}A_G^0=\Biggl[{{\Delta P_{Gq} \otimes \Delta u_v} \over {P_{Gq}
\otimes u_v}}\Biggr]. \label{gpr:2.5}
\end{equation}
in terms of the flavor non-singlet quark distributions, valid for large $x$.
We can then define the interative approximation:
\begin{equation}
A_G^{n+1}=\Biggl[{{\Delta P_{Gq} \otimes \Delta q+\Delta P_{GG}\otimes
(A_G^n\cdot G)} \over {P_{Gq} \otimes q+P_{GG}\otimes G}}\Biggr],
\label{2.6}
\end{equation}
which should converge for large enough n. 
For the starting distributions and the iterations, we use the polarized quark 
distributions outlined by GGR\cite{ggr} and the CTEQ4M unpolarized 
distributions.\cite{cteq} The iterations continue until a suitable convergence
is reached. The spin-weighted gluon asymmetry is then determined explicitly by 
$\Delta G(x,t)=A_G(x)\cdot G(x,t)$. The resulting shape of $A_G(x)$ is shown 
in Figure 1. This implies a larger polarized gluon distribution than the $xG$ 
model of GGRA\cite{ggr}, so spin asymmetries which depend upon $\Delta G$ are 
enhanced.

At small-$x$, the argument that ${{\partial A_G}\over {\partial t}}=0$
implies that $A_G$ maintains its $x$-dependent shape asymptotically in $t$. 
Discussion can be found in ref. [19]. The uncertainties at small-$x$, range
from $\pm 0.02$ at $x\approx 0.20$ up to $\pm 0.1$ below $x=0.05$. These are
primarily due to NLO corrections.

\begin{figure}
\begin{center}
\input{AG.tex}
\end{center}
\caption{The gluon asymmetry $\frac{\Delta G}{G}$ plotted as a function of
$x$}
\end{figure}
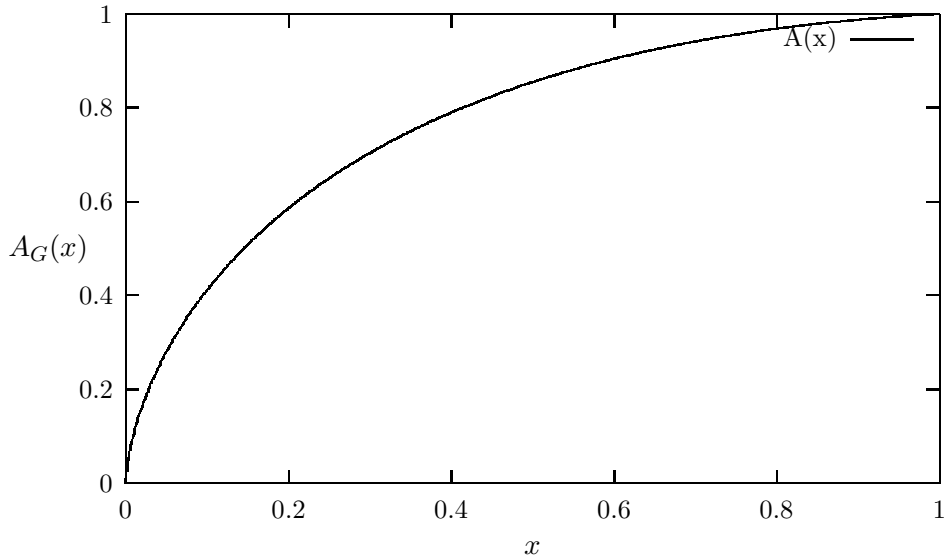

\section{Experimental Program}

There are many experiments either in progress or that have been proposed, to
provide the means to determine the polarized distributions.
These include:
\begin{itemize}
\item polarized lepton-hadron scattering (DIS)
\item polarized lepton pair production 
\item jet production 
\item direct photon production 
\item $\Delta G/G$ measurements 
\item cc events in $W^{\pm}$ production 
\item fragmentation in $\pi$ and $K$ production. 
\end{itemize}

Although polarized DIS does not provide complete information on the flavor
dependence of the polarized sea distributions, much can be learned from
precision $x$-dependent measurements. These wil provide a consistency check
for the separate distributions as they are determined. As discussed in ref. 
[22], polarized jet production and prompt photon production are the best
candidates for the determination of $\Delta G$. The appropriate kinematic
regions for STAR and PHENIX to search are also discussed in that paper.
There are a number of existing and planned experiments are suitable for
measuring either $A_G(x)$ or a combination of $\Delta G(x,Q^2)$ and $G(x,Q^2)$. 
Since this model of $\frac{\Delta G}{G}$ implies a larger polarized glue than 
the GGRA model used in ref. [22], all of the asymmetries for direct-$\gamma$ 
and jet production should be enhanced, making them easier to distinguish from 
parametrizations of $\Delta G$.

The HERMES experimental group at DESY has measured the longitudinal cross
section asymmetry $A_{\|}$ in high-$p_T$ hadronic photoproduction.\cite{hermes}
From this and known values of $\frac{\Delta q}{q}$ from DIS, a value for
$A_G(x_G)$ can be extracted. Here, $x_G=\hat{s}/2M\nu$ is the nucleon momentum
fraction carried by the gluon. Our value at $x_G=0.17$ is within one $\sigma$ 
of the quoted value of $A_G=0.41\pm 0.18$ (stat.) $\pm 0.03$ (syst.). 

The COMPASS group at CERN.\cite{compass} plans to extract $A_G$ from the photon
nucleon asymmetry, $A_{\gamma N}^{c\bar{c}}(x_G)$ in open charm muo-production,
which is dominated by the photon-gluon fusion process. This experiment should
be able to cover a wide kinematic range of $x_G$ as a further check of this
model. The combination of these experiments will be a good test of the
assumptions of our gluon asymmetry model and a consistency check on our
knowledge of the gluon polarization in the nucleon.

\section*{Acknowledgements}

The work on the polarized gluon asymmetry was done in collaboration with
F. Close and D. Sivers. Work supported in part by the U.S. Department of 
Energy, Division of High Energy Physics, Contract W-31-109-ENG-38.

\end{document}

%% file: AG.tex
\setlength{\unitlength}{0.240900pt}
\ifx\plotpoint\undefined\newsavebox{\plotpoint}\fi
\sbox{\plotpoint}{\rule[-0.200pt]{0.400pt}{0.400pt}}%
\begin{picture}(1500,900)(0,0)
\font\gnuplot=cmr10 at 10pt
\gnuplot
\sbox{\plotpoint}{\rule[-0.200pt]{0.400pt}{0.400pt}}%
\put(161.0,123.0){\rule[-0.200pt]{4.818pt}{0.400pt}}
\put(141,123){\makebox(0,0)[r]{0}}
\put(1419.0,123.0){\rule[-0.200pt]{4.818pt}{0.400pt}}
\put(161.0,270.0){\rule[-0.200pt]{4.818pt}{0.400pt}}
\put(141,270){\makebox(0,0)[r]{0.2}}
\put(1419.0,270.0){\rule[-0.200pt]{4.818pt}{0.400pt}}
\put(161.0,418.0){\rule[-0.200pt]{4.818pt}{0.400pt}}
\put(141,418){\makebox(0,0)[r]{0.4}}
\put(1419.0,418.0){\rule[-0.200pt]{4.818pt}{0.400pt}}
\put(161.0,565.0){\rule[-0.200pt]{4.818pt}{0.400pt}}
\put(141,565){\makebox(0,0)[r]{0.6}}
\put(1419.0,565.0){\rule[-0.200pt]{4.818pt}{0.400pt}}
\put(161.0,713.0){\rule[-0.200pt]{4.818pt}{0.400pt}}
\put(141,713){\makebox(0,0)[r]{0.8}}
\put(1419.0,713.0){\rule[-0.200pt]{4.818pt}{0.400pt}}
\put(161.0,860.0){\rule[-0.200pt]{4.818pt}{0.400pt}}
\put(141,860){\makebox(0,0)[r]{1}}
\put(1419.0,860.0){\rule[-0.200pt]{4.818pt}{0.400pt}}
\put(161.0,123.0){\rule[-0.200pt]{0.400pt}{4.818pt}}
\put(161,82){\makebox(0,0){0}}
\put(161.0,840.0){\rule[-0.200pt]{0.400pt}{4.818pt}}
\put(417.0,123.0){\rule[-0.200pt]{0.400pt}{4.818pt}}
\put(417,82){\makebox(0,0){0.2}}
\put(417.0,840.0){\rule[-0.200pt]{0.400pt}{4.818pt}}
\put(672.0,123.0){\rule[-0.200pt]{0.400pt}{4.818pt}}
\put(672,82){\makebox(0,0){0.4}}
\put(672.0,840.0){\rule[-0.200pt]{0.400pt}{4.818pt}}
\put(928.0,123.0){\rule[-0.200pt]{0.400pt}{4.818pt}}
\put(928,82){\makebox(0,0){0.6}}
\put(928.0,840.0){\rule[-0.200pt]{0.400pt}{4.818pt}}
\put(1183.0,123.0){\rule[-0.200pt]{0.400pt}{4.818pt}}
\put(1183,82){\makebox(0,0){0.8}}
\put(1183.0,840.0){\rule[-0.200pt]{0.400pt}{4.818pt}}
\put(1439.0,123.0){\rule[-0.200pt]{0.400pt}{4.818pt}}
\put(1439,82){\makebox(0,0){1}}
\put(1439.0,840.0){\rule[-0.200pt]{0.400pt}{4.818pt}}
\put(161.0,123.0){\rule[-0.200pt]{307.870pt}{0.400pt}}
\put(1439.0,123.0){\rule[-0.200pt]{0.400pt}{177.543pt}}
\put(161.0,860.0){\rule[-0.200pt]{307.870pt}{0.400pt}}
\put(40,491){\makebox(0,0){$A_G(x)$}}
\put(800,21){\makebox(0,0){$x$}}
\put(161.0,123.0){\rule[-0.200pt]{0.400pt}{177.543pt}}
\put(1279,820){\makebox(0,0)[r]{A(x)}}
\put(1299.0,820.0){\rule[-0.200pt]{24.090pt}{0.400pt}}
\put(161,123){\usebox{\plotpoint}}
\multiput(161.61,123.00)(0.447,5.597){3}{\rule{0.108pt}{3.567pt}}
\multiput(160.17,123.00)(3.000,18.597){2}{\rule{0.400pt}{1.783pt}}
\put(164.17,149){\rule{0.400pt}{3.300pt}}
\multiput(163.17,149.00)(2.000,9.151){2}{\rule{0.400pt}{1.650pt}}
\multiput(166.61,165.00)(0.447,2.695){3}{\rule{0.108pt}{1.833pt}}
\multiput(165.17,165.00)(3.000,9.195){2}{\rule{0.400pt}{0.917pt}}
\put(169.17,178){\rule{0.400pt}{2.300pt}}
\multiput(168.17,178.00)(2.000,6.226){2}{\rule{0.400pt}{1.150pt}}
\multiput(171.61,189.00)(0.447,2.025){3}{\rule{0.108pt}{1.433pt}}
\multiput(170.17,189.00)(3.000,7.025){2}{\rule{0.400pt}{0.717pt}}
\put(174.17,199){\rule{0.400pt}{2.100pt}}
\multiput(173.17,199.00)(2.000,5.641){2}{\rule{0.400pt}{1.050pt}}
\multiput(176.61,209.00)(0.447,1.802){3}{\rule{0.108pt}{1.300pt}}
\multiput(175.17,209.00)(3.000,6.302){2}{\rule{0.400pt}{0.650pt}}
\put(179.17,218){\rule{0.400pt}{1.700pt}}
\multiput(178.17,218.00)(2.000,4.472){2}{\rule{0.400pt}{0.850pt}}
\multiput(181.61,226.00)(0.447,1.579){3}{\rule{0.108pt}{1.167pt}}
\multiput(180.17,226.00)(3.000,5.579){2}{\rule{0.400pt}{0.583pt}}
\multiput(184.61,234.00)(0.447,1.355){3}{\rule{0.108pt}{1.033pt}}
\multiput(183.17,234.00)(3.000,4.855){2}{\rule{0.400pt}{0.517pt}}
\put(187.17,241){\rule{0.400pt}{1.500pt}}
\multiput(186.17,241.00)(2.000,3.887){2}{\rule{0.400pt}{0.750pt}}
\multiput(189.61,248.00)(0.447,1.355){3}{\rule{0.108pt}{1.033pt}}
\multiput(188.17,248.00)(3.000,4.855){2}{\rule{0.400pt}{0.517pt}}
\put(192.17,255){\rule{0.400pt}{1.500pt}}
\multiput(191.17,255.00)(2.000,3.887){2}{\rule{0.400pt}{0.750pt}}
\multiput(194.61,262.00)(0.447,1.132){3}{\rule{0.108pt}{0.900pt}}
\multiput(193.17,262.00)(3.000,4.132){2}{\rule{0.400pt}{0.450pt}}
\put(197.17,268){\rule{0.400pt}{1.500pt}}
\multiput(196.17,268.00)(2.000,3.887){2}{\rule{0.400pt}{0.750pt}}
\multiput(199.61,275.00)(0.447,1.132){3}{\rule{0.108pt}{0.900pt}}
\multiput(198.17,275.00)(3.000,4.132){2}{\rule{0.400pt}{0.450pt}}
\multiput(202.61,281.00)(0.447,1.132){3}{\rule{0.108pt}{0.900pt}}
\multiput(201.17,281.00)(3.000,4.132){2}{\rule{0.400pt}{0.450pt}}
\put(205.17,287){\rule{0.400pt}{1.100pt}}
\multiput(204.17,287.00)(2.000,2.717){2}{\rule{0.400pt}{0.550pt}}
\multiput(207.61,292.00)(0.447,1.132){3}{\rule{0.108pt}{0.900pt}}
\multiput(206.17,292.00)(3.000,4.132){2}{\rule{0.400pt}{0.450pt}}
\put(210.17,298){\rule{0.400pt}{1.100pt}}
\multiput(209.17,298.00)(2.000,2.717){2}{\rule{0.400pt}{0.550pt}}
\multiput(212.61,303.00)(0.447,0.909){3}{\rule{0.108pt}{0.767pt}}
\multiput(211.17,303.00)(3.000,3.409){2}{\rule{0.400pt}{0.383pt}}
\put(215.17,308){\rule{0.400pt}{1.300pt}}
\multiput(214.17,308.00)(2.000,3.302){2}{\rule{0.400pt}{0.650pt}}
\multiput(217.61,314.00)(0.447,0.909){3}{\rule{0.108pt}{0.767pt}}
\multiput(216.17,314.00)(3.000,3.409){2}{\rule{0.400pt}{0.383pt}}
\put(220.17,319){\rule{0.400pt}{1.100pt}}
\multiput(219.17,319.00)(2.000,2.717){2}{\rule{0.400pt}{0.550pt}}
\multiput(222.61,324.00)(0.447,0.685){3}{\rule{0.108pt}{0.633pt}}
\multiput(221.17,324.00)(3.000,2.685){2}{\rule{0.400pt}{0.317pt}}
\multiput(225.61,328.00)(0.447,0.909){3}{\rule{0.108pt}{0.767pt}}
\multiput(224.17,328.00)(3.000,3.409){2}{\rule{0.400pt}{0.383pt}}
\put(228.17,333){\rule{0.400pt}{1.100pt}}
\multiput(227.17,333.00)(2.000,2.717){2}{\rule{0.400pt}{0.550pt}}
\multiput(230.61,338.00)(0.447,0.685){3}{\rule{0.108pt}{0.633pt}}
\multiput(229.17,338.00)(3.000,2.685){2}{\rule{0.400pt}{0.317pt}}
\put(233.17,342){\rule{0.400pt}{1.100pt}}
\multiput(232.17,342.00)(2.000,2.717){2}{\rule{0.400pt}{0.550pt}}
\multiput(235.61,347.00)(0.447,0.685){3}{\rule{0.108pt}{0.633pt}}
\multiput(234.17,347.00)(3.000,2.685){2}{\rule{0.400pt}{0.317pt}}
\put(238.17,351){\rule{0.400pt}{0.900pt}}
\multiput(237.17,351.00)(2.000,2.132){2}{\rule{0.400pt}{0.450pt}}
\multiput(240.61,355.00)(0.447,0.909){3}{\rule{0.108pt}{0.767pt}}
\multiput(239.17,355.00)(3.000,3.409){2}{\rule{0.400pt}{0.383pt}}
\multiput(243.61,360.00)(0.447,0.685){3}{\rule{0.108pt}{0.633pt}}
\multiput(242.17,360.00)(3.000,2.685){2}{\rule{0.400pt}{0.317pt}}
\put(246.17,364){\rule{0.400pt}{0.900pt}}
\multiput(245.17,364.00)(2.000,2.132){2}{\rule{0.400pt}{0.450pt}}
\multiput(248.61,368.00)(0.447,0.685){3}{\rule{0.108pt}{0.633pt}}
\multiput(247.17,368.00)(3.000,2.685){2}{\rule{0.400pt}{0.317pt}}
\put(251.17,372){\rule{0.400pt}{0.900pt}}
\multiput(250.17,372.00)(2.000,2.132){2}{\rule{0.400pt}{0.450pt}}
\multiput(253.61,376.00)(0.447,0.685){3}{\rule{0.108pt}{0.633pt}}
\multiput(252.17,376.00)(3.000,2.685){2}{\rule{0.400pt}{0.317pt}}
\put(256.17,380){\rule{0.400pt}{0.900pt}}
\multiput(255.17,380.00)(2.000,2.132){2}{\rule{0.400pt}{0.450pt}}
\multiput(258.61,384.00)(0.447,0.685){3}{\rule{0.108pt}{0.633pt}}
\multiput(257.17,384.00)(3.000,2.685){2}{\rule{0.400pt}{0.317pt}}
\put(261.17,388){\rule{0.400pt}{0.700pt}}
\multiput(260.17,388.00)(2.000,1.547){2}{\rule{0.400pt}{0.350pt}}
\multiput(263.61,391.00)(0.447,0.685){3}{\rule{0.108pt}{0.633pt}}
\multiput(262.17,391.00)(3.000,2.685){2}{\rule{0.400pt}{0.317pt}}
\multiput(266.61,395.00)(0.447,0.685){3}{\rule{0.108pt}{0.633pt}}
\multiput(265.17,395.00)(3.000,2.685){2}{\rule{0.400pt}{0.317pt}}
\put(269.17,399){\rule{0.400pt}{0.700pt}}
\multiput(268.17,399.00)(2.000,1.547){2}{\rule{0.400pt}{0.350pt}}
\multiput(271.61,402.00)(0.447,0.685){3}{\rule{0.108pt}{0.633pt}}
\multiput(270.17,402.00)(3.000,2.685){2}{\rule{0.400pt}{0.317pt}}
\put(274.17,406){\rule{0.400pt}{0.700pt}}
\multiput(273.17,406.00)(2.000,1.547){2}{\rule{0.400pt}{0.350pt}}
\multiput(276.61,409.00)(0.447,0.685){3}{\rule{0.108pt}{0.633pt}}
\multiput(275.17,409.00)(3.000,2.685){2}{\rule{0.400pt}{0.317pt}}
\put(279.17,413){\rule{0.400pt}{0.700pt}}
\multiput(278.17,413.00)(2.000,1.547){2}{\rule{0.400pt}{0.350pt}}
\multiput(281.61,416.00)(0.447,0.685){3}{\rule{0.108pt}{0.633pt}}
\multiput(280.17,416.00)(3.000,2.685){2}{\rule{0.400pt}{0.317pt}}
\put(284.17,420){\rule{0.400pt}{0.700pt}}
\multiput(283.17,420.00)(2.000,1.547){2}{\rule{0.400pt}{0.350pt}}
\multiput(286.00,423.61)(0.462,0.447){3}{\rule{0.500pt}{0.108pt}}
\multiput(286.00,422.17)(1.962,3.000){2}{\rule{0.250pt}{0.400pt}}
\multiput(289.00,426.61)(0.462,0.447){3}{\rule{0.500pt}{0.108pt}}
\multiput(289.00,425.17)(1.962,3.000){2}{\rule{0.250pt}{0.400pt}}
\put(292.17,429){\rule{0.400pt}{0.900pt}}
\multiput(291.17,429.00)(2.000,2.132){2}{\rule{0.400pt}{0.450pt}}
\multiput(294.00,433.61)(0.462,0.447){3}{\rule{0.500pt}{0.108pt}}
\multiput(294.00,432.17)(1.962,3.000){2}{\rule{0.250pt}{0.400pt}}
\put(297.17,436){\rule{0.400pt}{0.700pt}}
\multiput(296.17,436.00)(2.000,1.547){2}{\rule{0.400pt}{0.350pt}}
\multiput(299.00,439.61)(0.462,0.447){3}{\rule{0.500pt}{0.108pt}}
\multiput(299.00,438.17)(1.962,3.000){2}{\rule{0.250pt}{0.400pt}}
\put(302.17,442){\rule{0.400pt}{0.700pt}}
\multiput(301.17,442.00)(2.000,1.547){2}{\rule{0.400pt}{0.350pt}}
\multiput(304.00,445.61)(0.462,0.447){3}{\rule{0.500pt}{0.108pt}}
\multiput(304.00,444.17)(1.962,3.000){2}{\rule{0.250pt}{0.400pt}}
\multiput(307.00,448.61)(0.462,0.447){3}{\rule{0.500pt}{0.108pt}}
\multiput(307.00,447.17)(1.962,3.000){2}{\rule{0.250pt}{0.400pt}}
\put(310.17,451){\rule{0.400pt}{0.700pt}}
\multiput(309.17,451.00)(2.000,1.547){2}{\rule{0.400pt}{0.350pt}}
\multiput(312.00,454.61)(0.462,0.447){3}{\rule{0.500pt}{0.108pt}}
\multiput(312.00,453.17)(1.962,3.000){2}{\rule{0.250pt}{0.400pt}}
\put(315.17,457){\rule{0.400pt}{0.700pt}}
\multiput(314.17,457.00)(2.000,1.547){2}{\rule{0.400pt}{0.350pt}}
\multiput(317.00,460.61)(0.462,0.447){3}{\rule{0.500pt}{0.108pt}}
\multiput(317.00,459.17)(1.962,3.000){2}{\rule{0.250pt}{0.400pt}}
\put(320.17,463){\rule{0.400pt}{0.700pt}}
\multiput(319.17,463.00)(2.000,1.547){2}{\rule{0.400pt}{0.350pt}}
\multiput(322.00,466.61)(0.462,0.447){3}{\rule{0.500pt}{0.108pt}}
\multiput(322.00,465.17)(1.962,3.000){2}{\rule{0.250pt}{0.400pt}}
\put(325,469.17){\rule{0.482pt}{0.400pt}}
\multiput(325.00,468.17)(1.000,2.000){2}{\rule{0.241pt}{0.400pt}}
\multiput(327.00,471.61)(0.462,0.447){3}{\rule{0.500pt}{0.108pt}}
\multiput(327.00,470.17)(1.962,3.000){2}{\rule{0.250pt}{0.400pt}}
\multiput(330.00,474.61)(0.462,0.447){3}{\rule{0.500pt}{0.108pt}}
\multiput(330.00,473.17)(1.962,3.000){2}{\rule{0.250pt}{0.400pt}}
\put(333.17,477){\rule{0.400pt}{0.700pt}}
\multiput(332.17,477.00)(2.000,1.547){2}{\rule{0.400pt}{0.350pt}}
\put(335,480.17){\rule{0.700pt}{0.400pt}}
\multiput(335.00,479.17)(1.547,2.000){2}{\rule{0.350pt}{0.400pt}}
\put(338.17,482){\rule{0.400pt}{0.700pt}}
\multiput(337.17,482.00)(2.000,1.547){2}{\rule{0.400pt}{0.350pt}}
\multiput(340.00,485.61)(0.462,0.447){3}{\rule{0.500pt}{0.108pt}}
\multiput(340.00,484.17)(1.962,3.000){2}{\rule{0.250pt}{0.400pt}}
\put(343,488.17){\rule{0.482pt}{0.400pt}}
\multiput(343.00,487.17)(1.000,2.000){2}{\rule{0.241pt}{0.400pt}}
\multiput(345.00,490.61)(0.462,0.447){3}{\rule{0.500pt}{0.108pt}}
\multiput(345.00,489.17)(1.962,3.000){2}{\rule{0.250pt}{0.400pt}}
\put(348,493.17){\rule{0.700pt}{0.400pt}}
\multiput(348.00,492.17)(1.547,2.000){2}{\rule{0.350pt}{0.400pt}}
\put(351.17,495){\rule{0.400pt}{0.700pt}}
\multiput(350.17,495.00)(2.000,1.547){2}{\rule{0.400pt}{0.350pt}}
\put(353,498.17){\rule{0.700pt}{0.400pt}}
\multiput(353.00,497.17)(1.547,2.000){2}{\rule{0.350pt}{0.400pt}}
\put(356.17,500){\rule{0.400pt}{0.700pt}}
\multiput(355.17,500.00)(2.000,1.547){2}{\rule{0.400pt}{0.350pt}}
\put(358,503.17){\rule{0.700pt}{0.400pt}}
\multiput(358.00,502.17)(1.547,2.000){2}{\rule{0.350pt}{0.400pt}}
\put(361.17,505){\rule{0.400pt}{0.700pt}}
\multiput(360.17,505.00)(2.000,1.547){2}{\rule{0.400pt}{0.350pt}}
\put(363,508.17){\rule{0.700pt}{0.400pt}}
\multiput(363.00,507.17)(1.547,2.000){2}{\rule{0.350pt}{0.400pt}}
\put(366.17,510){\rule{0.400pt}{0.700pt}}
\multiput(365.17,510.00)(2.000,1.547){2}{\rule{0.400pt}{0.350pt}}
\put(368,513.17){\rule{0.700pt}{0.400pt}}
\multiput(368.00,512.17)(1.547,2.000){2}{\rule{0.350pt}{0.400pt}}
\put(371,515.17){\rule{0.700pt}{0.400pt}}
\multiput(371.00,514.17)(1.547,2.000){2}{\rule{0.350pt}{0.400pt}}
\put(374.17,517){\rule{0.400pt}{0.700pt}}
\multiput(373.17,517.00)(2.000,1.547){2}{\rule{0.400pt}{0.350pt}}
\put(376,520.17){\rule{0.700pt}{0.400pt}}
\multiput(376.00,519.17)(1.547,2.000){2}{\rule{0.350pt}{0.400pt}}
\put(379,522.17){\rule{0.482pt}{0.400pt}}
\multiput(379.00,521.17)(1.000,2.000){2}{\rule{0.241pt}{0.400pt}}
\multiput(381.00,524.61)(0.462,0.447){3}{\rule{0.500pt}{0.108pt}}
\multiput(381.00,523.17)(1.962,3.000){2}{\rule{0.250pt}{0.400pt}}
\put(384,527.17){\rule{0.482pt}{0.400pt}}
\multiput(384.00,526.17)(1.000,2.000){2}{\rule{0.241pt}{0.400pt}}
\put(386,529.17){\rule{0.700pt}{0.400pt}}
\multiput(386.00,528.17)(1.547,2.000){2}{\rule{0.350pt}{0.400pt}}
\multiput(389.00,531.61)(0.462,0.447){3}{\rule{0.500pt}{0.108pt}}
\multiput(389.00,530.17)(1.962,3.000){2}{\rule{0.250pt}{0.400pt}}
\put(392,534.17){\rule{0.482pt}{0.400pt}}
\multiput(392.00,533.17)(1.000,2.000){2}{\rule{0.241pt}{0.400pt}}
\put(394,536.17){\rule{0.700pt}{0.400pt}}
\multiput(394.00,535.17)(1.547,2.000){2}{\rule{0.350pt}{0.400pt}}
\put(397,538.17){\rule{0.482pt}{0.400pt}}
\multiput(397.00,537.17)(1.000,2.000){2}{\rule{0.241pt}{0.400pt}}
\put(399,540.17){\rule{0.700pt}{0.400pt}}
\multiput(399.00,539.17)(1.547,2.000){2}{\rule{0.350pt}{0.400pt}}
\put(402,542.17){\rule{0.482pt}{0.400pt}}
\multiput(402.00,541.17)(1.000,2.000){2}{\rule{0.241pt}{0.400pt}}
\multiput(404.00,544.61)(0.462,0.447){3}{\rule{0.500pt}{0.108pt}}
\multiput(404.00,543.17)(1.962,3.000){2}{\rule{0.250pt}{0.400pt}}
\put(407,547.17){\rule{0.482pt}{0.400pt}}
\multiput(407.00,546.17)(1.000,2.000){2}{\rule{0.241pt}{0.400pt}}
\put(409,549.17){\rule{0.700pt}{0.400pt}}
\multiput(409.00,548.17)(1.547,2.000){2}{\rule{0.350pt}{0.400pt}}
\put(412,551.17){\rule{0.700pt}{0.400pt}}
\multiput(412.00,550.17)(1.547,2.000){2}{\rule{0.350pt}{0.400pt}}
\put(415,553.17){\rule{0.482pt}{0.400pt}}
\multiput(415.00,552.17)(1.000,2.000){2}{\rule{0.241pt}{0.400pt}}
\put(417,555.17){\rule{0.700pt}{0.400pt}}
\multiput(417.00,554.17)(1.547,2.000){2}{\rule{0.350pt}{0.400pt}}
\put(420,557.17){\rule{0.482pt}{0.400pt}}
\multiput(420.00,556.17)(1.000,2.000){2}{\rule{0.241pt}{0.400pt}}
\put(422,559.17){\rule{0.700pt}{0.400pt}}
\multiput(422.00,558.17)(1.547,2.000){2}{\rule{0.350pt}{0.400pt}}
\put(425,561.17){\rule{0.482pt}{0.400pt}}
\multiput(425.00,560.17)(1.000,2.000){2}{\rule{0.241pt}{0.400pt}}
\put(427,563.17){\rule{0.700pt}{0.400pt}}
\multiput(427.00,562.17)(1.547,2.000){2}{\rule{0.350pt}{0.400pt}}
\put(430,565.17){\rule{0.482pt}{0.400pt}}
\multiput(430.00,564.17)(1.000,2.000){2}{\rule{0.241pt}{0.400pt}}
\put(432,567.17){\rule{0.700pt}{0.400pt}}
\multiput(432.00,566.17)(1.547,2.000){2}{\rule{0.350pt}{0.400pt}}
\put(435,569.17){\rule{0.700pt}{0.400pt}}
\multiput(435.00,568.17)(1.547,2.000){2}{\rule{0.350pt}{0.400pt}}
\put(438,571.17){\rule{0.482pt}{0.400pt}}
\multiput(438.00,570.17)(1.000,2.000){2}{\rule{0.241pt}{0.400pt}}
\put(440,573.17){\rule{0.700pt}{0.400pt}}
\multiput(440.00,572.17)(1.547,2.000){2}{\rule{0.350pt}{0.400pt}}
\put(443,575.17){\rule{0.482pt}{0.400pt}}
\multiput(443.00,574.17)(1.000,2.000){2}{\rule{0.241pt}{0.400pt}}
\put(445,577.17){\rule{0.700pt}{0.400pt}}
\multiput(445.00,576.17)(1.547,2.000){2}{\rule{0.350pt}{0.400pt}}
\put(448,579.17){\rule{0.482pt}{0.400pt}}
\multiput(448.00,578.17)(1.000,2.000){2}{\rule{0.241pt}{0.400pt}}
\put(450,580.67){\rule{0.723pt}{0.400pt}}
\multiput(450.00,580.17)(1.500,1.000){2}{\rule{0.361pt}{0.400pt}}
\put(453,582.17){\rule{0.700pt}{0.400pt}}
\multiput(453.00,581.17)(1.547,2.000){2}{\rule{0.350pt}{0.400pt}}
\put(456,584.17){\rule{0.482pt}{0.400pt}}
\multiput(456.00,583.17)(1.000,2.000){2}{\rule{0.241pt}{0.400pt}}
\put(458,586.17){\rule{0.700pt}{0.400pt}}
\multiput(458.00,585.17)(1.547,2.000){2}{\rule{0.350pt}{0.400pt}}
\put(461,588.17){\rule{0.482pt}{0.400pt}}
\multiput(461.00,587.17)(1.000,2.000){2}{\rule{0.241pt}{0.400pt}}
\put(463,590.17){\rule{0.700pt}{0.400pt}}
\multiput(463.00,589.17)(1.547,2.000){2}{\rule{0.350pt}{0.400pt}}
\put(466,591.67){\rule{0.482pt}{0.400pt}}
\multiput(466.00,591.17)(1.000,1.000){2}{\rule{0.241pt}{0.400pt}}
\put(468,593.17){\rule{0.700pt}{0.400pt}}
\multiput(468.00,592.17)(1.547,2.000){2}{\rule{0.350pt}{0.400pt}}
\put(471,595.17){\rule{0.482pt}{0.400pt}}
\multiput(471.00,594.17)(1.000,2.000){2}{\rule{0.241pt}{0.400pt}}
\put(473,597.17){\rule{0.700pt}{0.400pt}}
\multiput(473.00,596.17)(1.547,2.000){2}{\rule{0.350pt}{0.400pt}}
\put(476,598.67){\rule{0.723pt}{0.400pt}}
\multiput(476.00,598.17)(1.500,1.000){2}{\rule{0.361pt}{0.400pt}}
\put(479,600.17){\rule{0.482pt}{0.400pt}}
\multiput(479.00,599.17)(1.000,2.000){2}{\rule{0.241pt}{0.400pt}}
\put(481,602.17){\rule{0.700pt}{0.400pt}}
\multiput(481.00,601.17)(1.547,2.000){2}{\rule{0.350pt}{0.400pt}}
\put(484,604.17){\rule{0.482pt}{0.400pt}}
\multiput(484.00,603.17)(1.000,2.000){2}{\rule{0.241pt}{0.400pt}}
\put(486,605.67){\rule{0.723pt}{0.400pt}}
\multiput(486.00,605.17)(1.500,1.000){2}{\rule{0.361pt}{0.400pt}}
\put(489,607.17){\rule{0.482pt}{0.400pt}}
\multiput(489.00,606.17)(1.000,2.000){2}{\rule{0.241pt}{0.400pt}}
\put(491,609.17){\rule{0.700pt}{0.400pt}}
\multiput(491.00,608.17)(1.547,2.000){2}{\rule{0.350pt}{0.400pt}}
\put(494,610.67){\rule{0.723pt}{0.400pt}}
\multiput(494.00,610.17)(1.500,1.000){2}{\rule{0.361pt}{0.400pt}}
\put(497,612.17){\rule{0.482pt}{0.400pt}}
\multiput(497.00,611.17)(1.000,2.000){2}{\rule{0.241pt}{0.400pt}}
\put(499,614.17){\rule{0.700pt}{0.400pt}}
\multiput(499.00,613.17)(1.547,2.000){2}{\rule{0.350pt}{0.400pt}}
\put(502,615.67){\rule{0.482pt}{0.400pt}}
\multiput(502.00,615.17)(1.000,1.000){2}{\rule{0.241pt}{0.400pt}}
\put(504,617.17){\rule{0.700pt}{0.400pt}}
\multiput(504.00,616.17)(1.547,2.000){2}{\rule{0.350pt}{0.400pt}}
\put(507,618.67){\rule{0.482pt}{0.400pt}}
\multiput(507.00,618.17)(1.000,1.000){2}{\rule{0.241pt}{0.400pt}}
\put(509,620.17){\rule{0.700pt}{0.400pt}}
\multiput(509.00,619.17)(1.547,2.000){2}{\rule{0.350pt}{0.400pt}}
\put(512,622.17){\rule{0.482pt}{0.400pt}}
\multiput(512.00,621.17)(1.000,2.000){2}{\rule{0.241pt}{0.400pt}}
\put(514,623.67){\rule{0.723pt}{0.400pt}}
\multiput(514.00,623.17)(1.500,1.000){2}{\rule{0.361pt}{0.400pt}}
\put(517,625.17){\rule{0.700pt}{0.400pt}}
\multiput(517.00,624.17)(1.547,2.000){2}{\rule{0.350pt}{0.400pt}}
\put(520,626.67){\rule{0.482pt}{0.400pt}}
\multiput(520.00,626.17)(1.000,1.000){2}{\rule{0.241pt}{0.400pt}}
\put(522,628.17){\rule{0.700pt}{0.400pt}}
\multiput(522.00,627.17)(1.547,2.000){2}{\rule{0.350pt}{0.400pt}}
\put(525,629.67){\rule{0.482pt}{0.400pt}}
\multiput(525.00,629.17)(1.000,1.000){2}{\rule{0.241pt}{0.400pt}}
\put(527,631.17){\rule{0.700pt}{0.400pt}}
\multiput(527.00,630.17)(1.547,2.000){2}{\rule{0.350pt}{0.400pt}}
\put(530,632.67){\rule{0.482pt}{0.400pt}}
\multiput(530.00,632.17)(1.000,1.000){2}{\rule{0.241pt}{0.400pt}}
\put(532,634.17){\rule{0.700pt}{0.400pt}}
\multiput(532.00,633.17)(1.547,2.000){2}{\rule{0.350pt}{0.400pt}}
\put(535,636.17){\rule{0.482pt}{0.400pt}}
\multiput(535.00,635.17)(1.000,2.000){2}{\rule{0.241pt}{0.400pt}}
\put(537,637.67){\rule{0.723pt}{0.400pt}}
\multiput(537.00,637.17)(1.500,1.000){2}{\rule{0.361pt}{0.400pt}}
\put(540,638.67){\rule{0.723pt}{0.400pt}}
\multiput(540.00,638.17)(1.500,1.000){2}{\rule{0.361pt}{0.400pt}}
\put(543,640.17){\rule{0.482pt}{0.400pt}}
\multiput(543.00,639.17)(1.000,2.000){2}{\rule{0.241pt}{0.400pt}}
\put(545,641.67){\rule{0.723pt}{0.400pt}}
\multiput(545.00,641.17)(1.500,1.000){2}{\rule{0.361pt}{0.400pt}}
\put(548,643.17){\rule{0.482pt}{0.400pt}}
\multiput(548.00,642.17)(1.000,2.000){2}{\rule{0.241pt}{0.400pt}}
\put(550,644.67){\rule{0.723pt}{0.400pt}}
\multiput(550.00,644.17)(1.500,1.000){2}{\rule{0.361pt}{0.400pt}}
\put(553,646.17){\rule{0.482pt}{0.400pt}}
\multiput(553.00,645.17)(1.000,2.000){2}{\rule{0.241pt}{0.400pt}}
\put(555,647.67){\rule{0.723pt}{0.400pt}}
\multiput(555.00,647.17)(1.500,1.000){2}{\rule{0.361pt}{0.400pt}}
\put(558,649.17){\rule{0.700pt}{0.400pt}}
\multiput(558.00,648.17)(1.547,2.000){2}{\rule{0.350pt}{0.400pt}}
\put(561,650.67){\rule{0.482pt}{0.400pt}}
\multiput(561.00,650.17)(1.000,1.000){2}{\rule{0.241pt}{0.400pt}}
\put(563,651.67){\rule{0.723pt}{0.400pt}}
\multiput(563.00,651.17)(1.500,1.000){2}{\rule{0.361pt}{0.400pt}}
\put(566,653.17){\rule{0.482pt}{0.400pt}}
\multiput(566.00,652.17)(1.000,2.000){2}{\rule{0.241pt}{0.400pt}}
\put(568,654.67){\rule{0.723pt}{0.400pt}}
\multiput(568.00,654.17)(1.500,1.000){2}{\rule{0.361pt}{0.400pt}}
\put(571,656.17){\rule{0.482pt}{0.400pt}}
\multiput(571.00,655.17)(1.000,2.000){2}{\rule{0.241pt}{0.400pt}}
\put(573,657.67){\rule{0.723pt}{0.400pt}}
\multiput(573.00,657.17)(1.500,1.000){2}{\rule{0.361pt}{0.400pt}}
\put(576,658.67){\rule{0.482pt}{0.400pt}}
\multiput(576.00,658.17)(1.000,1.000){2}{\rule{0.241pt}{0.400pt}}
\put(578,660.17){\rule{0.700pt}{0.400pt}}
\multiput(578.00,659.17)(1.547,2.000){2}{\rule{0.350pt}{0.400pt}}
\put(581,661.67){\rule{0.723pt}{0.400pt}}
\multiput(581.00,661.17)(1.500,1.000){2}{\rule{0.361pt}{0.400pt}}
\put(584,662.67){\rule{0.482pt}{0.400pt}}
\multiput(584.00,662.17)(1.000,1.000){2}{\rule{0.241pt}{0.400pt}}
\put(586,664.17){\rule{0.700pt}{0.400pt}}
\multiput(586.00,663.17)(1.547,2.000){2}{\rule{0.350pt}{0.400pt}}
\put(589,665.67){\rule{0.482pt}{0.400pt}}
\multiput(589.00,665.17)(1.000,1.000){2}{\rule{0.241pt}{0.400pt}}
\put(591,666.67){\rule{0.723pt}{0.400pt}}
\multiput(591.00,666.17)(1.500,1.000){2}{\rule{0.361pt}{0.400pt}}
\put(594,668.17){\rule{0.482pt}{0.400pt}}
\multiput(594.00,667.17)(1.000,2.000){2}{\rule{0.241pt}{0.400pt}}
\put(596,669.67){\rule{0.723pt}{0.400pt}}
\multiput(596.00,669.17)(1.500,1.000){2}{\rule{0.361pt}{0.400pt}}
\put(599,670.67){\rule{0.723pt}{0.400pt}}
\multiput(599.00,670.17)(1.500,1.000){2}{\rule{0.361pt}{0.400pt}}
\put(602,672.17){\rule{0.482pt}{0.400pt}}
\multiput(602.00,671.17)(1.000,2.000){2}{\rule{0.241pt}{0.400pt}}
\put(604,673.67){\rule{0.723pt}{0.400pt}}
\multiput(604.00,673.17)(1.500,1.000){2}{\rule{0.361pt}{0.400pt}}
\put(607,674.67){\rule{0.482pt}{0.400pt}}
\multiput(607.00,674.17)(1.000,1.000){2}{\rule{0.241pt}{0.400pt}}
\put(609,675.67){\rule{0.723pt}{0.400pt}}
\multiput(609.00,675.17)(1.500,1.000){2}{\rule{0.361pt}{0.400pt}}
\put(612,677.17){\rule{0.482pt}{0.400pt}}
\multiput(612.00,676.17)(1.000,2.000){2}{\rule{0.241pt}{0.400pt}}
\put(614,678.67){\rule{0.723pt}{0.400pt}}
\multiput(614.00,678.17)(1.500,1.000){2}{\rule{0.361pt}{0.400pt}}
\put(617,679.67){\rule{0.482pt}{0.400pt}}
\multiput(617.00,679.17)(1.000,1.000){2}{\rule{0.241pt}{0.400pt}}
\put(619,680.67){\rule{0.723pt}{0.400pt}}
\multiput(619.00,680.17)(1.500,1.000){2}{\rule{0.361pt}{0.400pt}}
\put(622,682.17){\rule{0.700pt}{0.400pt}}
\multiput(622.00,681.17)(1.547,2.000){2}{\rule{0.350pt}{0.400pt}}
\put(625,683.67){\rule{0.482pt}{0.400pt}}
\multiput(625.00,683.17)(1.000,1.000){2}{\rule{0.241pt}{0.400pt}}
\put(627,684.67){\rule{0.723pt}{0.400pt}}
\multiput(627.00,684.17)(1.500,1.000){2}{\rule{0.361pt}{0.400pt}}
\put(630,685.67){\rule{0.482pt}{0.400pt}}
\multiput(630.00,685.17)(1.000,1.000){2}{\rule{0.241pt}{0.400pt}}
\put(632,686.67){\rule{0.723pt}{0.400pt}}
\multiput(632.00,686.17)(1.500,1.000){2}{\rule{0.361pt}{0.400pt}}
\put(635,688.17){\rule{0.482pt}{0.400pt}}
\multiput(635.00,687.17)(1.000,2.000){2}{\rule{0.241pt}{0.400pt}}
\put(637,689.67){\rule{0.723pt}{0.400pt}}
\multiput(637.00,689.17)(1.500,1.000){2}{\rule{0.361pt}{0.400pt}}
\put(640,690.67){\rule{0.482pt}{0.400pt}}
\multiput(640.00,690.17)(1.000,1.000){2}{\rule{0.241pt}{0.400pt}}
\put(642,691.67){\rule{0.723pt}{0.400pt}}
\multiput(642.00,691.17)(1.500,1.000){2}{\rule{0.361pt}{0.400pt}}
\put(645,692.67){\rule{0.723pt}{0.400pt}}
\multiput(645.00,692.17)(1.500,1.000){2}{\rule{0.361pt}{0.400pt}}
\put(648,693.67){\rule{0.482pt}{0.400pt}}
\multiput(648.00,693.17)(1.000,1.000){2}{\rule{0.241pt}{0.400pt}}
\put(650,695.17){\rule{0.700pt}{0.400pt}}
\multiput(650.00,694.17)(1.547,2.000){2}{\rule{0.350pt}{0.400pt}}
\put(653,696.67){\rule{0.482pt}{0.400pt}}
\multiput(653.00,696.17)(1.000,1.000){2}{\rule{0.241pt}{0.400pt}}
\put(655,697.67){\rule{0.723pt}{0.400pt}}
\multiput(655.00,697.17)(1.500,1.000){2}{\rule{0.361pt}{0.400pt}}
\put(658,698.67){\rule{0.482pt}{0.400pt}}
\multiput(658.00,698.17)(1.000,1.000){2}{\rule{0.241pt}{0.400pt}}
\put(660,699.67){\rule{0.723pt}{0.400pt}}
\multiput(660.00,699.17)(1.500,1.000){2}{\rule{0.361pt}{0.400pt}}
\put(663,700.67){\rule{0.723pt}{0.400pt}}
\multiput(663.00,700.17)(1.500,1.000){2}{\rule{0.361pt}{0.400pt}}
\put(666,701.67){\rule{0.482pt}{0.400pt}}
\multiput(666.00,701.17)(1.000,1.000){2}{\rule{0.241pt}{0.400pt}}
\put(668,702.67){\rule{0.723pt}{0.400pt}}
\multiput(668.00,702.17)(1.500,1.000){2}{\rule{0.361pt}{0.400pt}}
\put(671,704.17){\rule{0.482pt}{0.400pt}}
\multiput(671.00,703.17)(1.000,2.000){2}{\rule{0.241pt}{0.400pt}}
\put(673,705.67){\rule{0.723pt}{0.400pt}}
\multiput(673.00,705.17)(1.500,1.000){2}{\rule{0.361pt}{0.400pt}}
\put(676,706.67){\rule{0.482pt}{0.400pt}}
\multiput(676.00,706.17)(1.000,1.000){2}{\rule{0.241pt}{0.400pt}}
\put(678,707.67){\rule{0.723pt}{0.400pt}}
\multiput(678.00,707.17)(1.500,1.000){2}{\rule{0.361pt}{0.400pt}}
\put(681,708.67){\rule{0.482pt}{0.400pt}}
\multiput(681.00,708.17)(1.000,1.000){2}{\rule{0.241pt}{0.400pt}}
\put(683,709.67){\rule{0.723pt}{0.400pt}}
\multiput(683.00,709.17)(1.500,1.000){2}{\rule{0.361pt}{0.400pt}}
\put(686,710.67){\rule{0.723pt}{0.400pt}}
\multiput(686.00,710.17)(1.500,1.000){2}{\rule{0.361pt}{0.400pt}}
\put(689,711.67){\rule{0.482pt}{0.400pt}}
\multiput(689.00,711.17)(1.000,1.000){2}{\rule{0.241pt}{0.400pt}}
\put(691,712.67){\rule{0.723pt}{0.400pt}}
\multiput(691.00,712.17)(1.500,1.000){2}{\rule{0.361pt}{0.400pt}}
\put(694,713.67){\rule{0.482pt}{0.400pt}}
\multiput(694.00,713.17)(1.000,1.000){2}{\rule{0.241pt}{0.400pt}}
\put(696,714.67){\rule{0.723pt}{0.400pt}}
\multiput(696.00,714.17)(1.500,1.000){2}{\rule{0.361pt}{0.400pt}}
\put(699,715.67){\rule{0.482pt}{0.400pt}}
\multiput(699.00,715.17)(1.000,1.000){2}{\rule{0.241pt}{0.400pt}}
\put(701,716.67){\rule{0.723pt}{0.400pt}}
\multiput(701.00,716.17)(1.500,1.000){2}{\rule{0.361pt}{0.400pt}}
\put(704,717.67){\rule{0.723pt}{0.400pt}}
\multiput(704.00,717.17)(1.500,1.000){2}{\rule{0.361pt}{0.400pt}}
\put(707,718.67){\rule{0.482pt}{0.400pt}}
\multiput(707.00,718.17)(1.000,1.000){2}{\rule{0.241pt}{0.400pt}}
\put(709,719.67){\rule{0.723pt}{0.400pt}}
\multiput(709.00,719.17)(1.500,1.000){2}{\rule{0.361pt}{0.400pt}}
\put(712,720.67){\rule{0.482pt}{0.400pt}}
\multiput(712.00,720.17)(1.000,1.000){2}{\rule{0.241pt}{0.400pt}}
\put(714,721.67){\rule{0.723pt}{0.400pt}}
\multiput(714.00,721.17)(1.500,1.000){2}{\rule{0.361pt}{0.400pt}}
\put(717,722.67){\rule{0.482pt}{0.400pt}}
\multiput(717.00,722.17)(1.000,1.000){2}{\rule{0.241pt}{0.400pt}}
\put(719,723.67){\rule{0.723pt}{0.400pt}}
\multiput(719.00,723.17)(1.500,1.000){2}{\rule{0.361pt}{0.400pt}}
\put(722,724.67){\rule{0.482pt}{0.400pt}}
\multiput(722.00,724.17)(1.000,1.000){2}{\rule{0.241pt}{0.400pt}}
\put(724,725.67){\rule{0.723pt}{0.400pt}}
\multiput(724.00,725.17)(1.500,1.000){2}{\rule{0.361pt}{0.400pt}}
\put(727,726.67){\rule{0.723pt}{0.400pt}}
\multiput(727.00,726.17)(1.500,1.000){2}{\rule{0.361pt}{0.400pt}}
\put(730,727.67){\rule{0.482pt}{0.400pt}}
\multiput(730.00,727.17)(1.000,1.000){2}{\rule{0.241pt}{0.400pt}}
\put(732,728.67){\rule{0.723pt}{0.400pt}}
\multiput(732.00,728.17)(1.500,1.000){2}{\rule{0.361pt}{0.400pt}}
\put(735,729.67){\rule{0.482pt}{0.400pt}}
\multiput(735.00,729.17)(1.000,1.000){2}{\rule{0.241pt}{0.400pt}}
\put(737,730.67){\rule{0.723pt}{0.400pt}}
\multiput(737.00,730.17)(1.500,1.000){2}{\rule{0.361pt}{0.400pt}}
\put(740,731.67){\rule{0.482pt}{0.400pt}}
\multiput(740.00,731.17)(1.000,1.000){2}{\rule{0.241pt}{0.400pt}}
\put(742,732.67){\rule{0.723pt}{0.400pt}}
\multiput(742.00,732.17)(1.500,1.000){2}{\rule{0.361pt}{0.400pt}}
\put(745,733.67){\rule{0.482pt}{0.400pt}}
\multiput(745.00,733.17)(1.000,1.000){2}{\rule{0.241pt}{0.400pt}}
\put(747,734.67){\rule{0.723pt}{0.400pt}}
\multiput(747.00,734.17)(1.500,1.000){2}{\rule{0.361pt}{0.400pt}}
\put(750,735.67){\rule{0.723pt}{0.400pt}}
\multiput(750.00,735.17)(1.500,1.000){2}{\rule{0.361pt}{0.400pt}}
\put(753,736.67){\rule{0.482pt}{0.400pt}}
\multiput(753.00,736.17)(1.000,1.000){2}{\rule{0.241pt}{0.400pt}}
\put(755,737.67){\rule{0.723pt}{0.400pt}}
\multiput(755.00,737.17)(1.500,1.000){2}{\rule{0.361pt}{0.400pt}}
\put(758,738.67){\rule{0.482pt}{0.400pt}}
\multiput(758.00,738.17)(1.000,1.000){2}{\rule{0.241pt}{0.400pt}}
\put(763,739.67){\rule{0.482pt}{0.400pt}}
\multiput(763.00,739.17)(1.000,1.000){2}{\rule{0.241pt}{0.400pt}}
\put(765,740.67){\rule{0.723pt}{0.400pt}}
\multiput(765.00,740.17)(1.500,1.000){2}{\rule{0.361pt}{0.400pt}}
\put(768,741.67){\rule{0.723pt}{0.400pt}}
\multiput(768.00,741.17)(1.500,1.000){2}{\rule{0.361pt}{0.400pt}}
\put(771,742.67){\rule{0.482pt}{0.400pt}}
\multiput(771.00,742.17)(1.000,1.000){2}{\rule{0.241pt}{0.400pt}}
\put(773,743.67){\rule{0.723pt}{0.400pt}}
\multiput(773.00,743.17)(1.500,1.000){2}{\rule{0.361pt}{0.400pt}}
\put(776,744.67){\rule{0.482pt}{0.400pt}}
\multiput(776.00,744.17)(1.000,1.000){2}{\rule{0.241pt}{0.400pt}}
\put(778,745.67){\rule{0.723pt}{0.400pt}}
\multiput(778.00,745.17)(1.500,1.000){2}{\rule{0.361pt}{0.400pt}}
\put(781,746.67){\rule{0.482pt}{0.400pt}}
\multiput(781.00,746.17)(1.000,1.000){2}{\rule{0.241pt}{0.400pt}}
\put(760.0,740.0){\rule[-0.200pt]{0.723pt}{0.400pt}}
\put(786,747.67){\rule{0.482pt}{0.400pt}}
\multiput(786.00,747.17)(1.000,1.000){2}{\rule{0.241pt}{0.400pt}}
\put(788,748.67){\rule{0.723pt}{0.400pt}}
\multiput(788.00,748.17)(1.500,1.000){2}{\rule{0.361pt}{0.400pt}}
\put(791,749.67){\rule{0.723pt}{0.400pt}}
\multiput(791.00,749.17)(1.500,1.000){2}{\rule{0.361pt}{0.400pt}}
\put(794,750.67){\rule{0.482pt}{0.400pt}}
\multiput(794.00,750.17)(1.000,1.000){2}{\rule{0.241pt}{0.400pt}}
\put(796,751.67){\rule{0.723pt}{0.400pt}}
\multiput(796.00,751.17)(1.500,1.000){2}{\rule{0.361pt}{0.400pt}}
\put(783.0,748.0){\rule[-0.200pt]{0.723pt}{0.400pt}}
\put(801,752.67){\rule{0.723pt}{0.400pt}}
\multiput(801.00,752.17)(1.500,1.000){2}{\rule{0.361pt}{0.400pt}}
\put(804,753.67){\rule{0.482pt}{0.400pt}}
\multiput(804.00,753.17)(1.000,1.000){2}{\rule{0.241pt}{0.400pt}}
\put(806,754.67){\rule{0.723pt}{0.400pt}}
\multiput(806.00,754.17)(1.500,1.000){2}{\rule{0.361pt}{0.400pt}}
\put(809,755.67){\rule{0.723pt}{0.400pt}}
\multiput(809.00,755.17)(1.500,1.000){2}{\rule{0.361pt}{0.400pt}}
\put(812,756.67){\rule{0.482pt}{0.400pt}}
\multiput(812.00,756.17)(1.000,1.000){2}{\rule{0.241pt}{0.400pt}}
\put(799.0,753.0){\rule[-0.200pt]{0.482pt}{0.400pt}}
\put(817,757.67){\rule{0.482pt}{0.400pt}}
\multiput(817.00,757.17)(1.000,1.000){2}{\rule{0.241pt}{0.400pt}}
\put(819,758.67){\rule{0.723pt}{0.400pt}}
\multiput(819.00,758.17)(1.500,1.000){2}{\rule{0.361pt}{0.400pt}}
\put(822,759.67){\rule{0.482pt}{0.400pt}}
\multiput(822.00,759.17)(1.000,1.000){2}{\rule{0.241pt}{0.400pt}}
\put(824,760.67){\rule{0.723pt}{0.400pt}}
\multiput(824.00,760.17)(1.500,1.000){2}{\rule{0.361pt}{0.400pt}}
\put(814.0,758.0){\rule[-0.200pt]{0.723pt}{0.400pt}}
\put(829,761.67){\rule{0.723pt}{0.400pt}}
\multiput(829.00,761.17)(1.500,1.000){2}{\rule{0.361pt}{0.400pt}}
\put(832,762.67){\rule{0.723pt}{0.400pt}}
\multiput(832.00,762.17)(1.500,1.000){2}{\rule{0.361pt}{0.400pt}}
\put(835,763.67){\rule{0.482pt}{0.400pt}}
\multiput(835.00,763.17)(1.000,1.000){2}{\rule{0.241pt}{0.400pt}}
\put(827.0,762.0){\rule[-0.200pt]{0.482pt}{0.400pt}}
\put(840,764.67){\rule{0.482pt}{0.400pt}}
\multiput(840.00,764.17)(1.000,1.000){2}{\rule{0.241pt}{0.400pt}}
\put(842,765.67){\rule{0.723pt}{0.400pt}}
\multiput(842.00,765.17)(1.500,1.000){2}{\rule{0.361pt}{0.400pt}}
\put(845,766.67){\rule{0.482pt}{0.400pt}}
\multiput(845.00,766.17)(1.000,1.000){2}{\rule{0.241pt}{0.400pt}}
\put(847,767.67){\rule{0.723pt}{0.400pt}}
\multiput(847.00,767.17)(1.500,1.000){2}{\rule{0.361pt}{0.400pt}}
\put(837.0,765.0){\rule[-0.200pt]{0.723pt}{0.400pt}}
\put(853,768.67){\rule{0.482pt}{0.400pt}}
\multiput(853.00,768.17)(1.000,1.000){2}{\rule{0.241pt}{0.400pt}}
\put(855,769.67){\rule{0.723pt}{0.400pt}}
\multiput(855.00,769.17)(1.500,1.000){2}{\rule{0.361pt}{0.400pt}}
\put(850.0,769.0){\rule[-0.200pt]{0.723pt}{0.400pt}}
\put(860,770.67){\rule{0.723pt}{0.400pt}}
\multiput(860.00,770.17)(1.500,1.000){2}{\rule{0.361pt}{0.400pt}}
\put(863,771.67){\rule{0.482pt}{0.400pt}}
\multiput(863.00,771.17)(1.000,1.000){2}{\rule{0.241pt}{0.400pt}}
\put(865,772.67){\rule{0.723pt}{0.400pt}}
\multiput(865.00,772.17)(1.500,1.000){2}{\rule{0.361pt}{0.400pt}}
\put(858.0,771.0){\rule[-0.200pt]{0.482pt}{0.400pt}}
\put(870,773.67){\rule{0.723pt}{0.400pt}}
\multiput(870.00,773.17)(1.500,1.000){2}{\rule{0.361pt}{0.400pt}}
\put(873,774.67){\rule{0.723pt}{0.400pt}}
\multiput(873.00,774.17)(1.500,1.000){2}{\rule{0.361pt}{0.400pt}}
\put(876,775.67){\rule{0.482pt}{0.400pt}}
\multiput(876.00,775.17)(1.000,1.000){2}{\rule{0.241pt}{0.400pt}}
\put(868.0,774.0){\rule[-0.200pt]{0.482pt}{0.400pt}}
\put(881,776.67){\rule{0.482pt}{0.400pt}}
\multiput(881.00,776.17)(1.000,1.000){2}{\rule{0.241pt}{0.400pt}}
\put(883,777.67){\rule{0.723pt}{0.400pt}}
\multiput(883.00,777.17)(1.500,1.000){2}{\rule{0.361pt}{0.400pt}}
\put(878.0,777.0){\rule[-0.200pt]{0.723pt}{0.400pt}}
\put(888,778.67){\rule{0.723pt}{0.400pt}}
\multiput(888.00,778.17)(1.500,1.000){2}{\rule{0.361pt}{0.400pt}}
\put(891,779.67){\rule{0.482pt}{0.400pt}}
\multiput(891.00,779.17)(1.000,1.000){2}{\rule{0.241pt}{0.400pt}}
\put(886.0,779.0){\rule[-0.200pt]{0.482pt}{0.400pt}}
\put(896,780.67){\rule{0.723pt}{0.400pt}}
\multiput(896.00,780.17)(1.500,1.000){2}{\rule{0.361pt}{0.400pt}}
\put(899,781.67){\rule{0.482pt}{0.400pt}}
\multiput(899.00,781.17)(1.000,1.000){2}{\rule{0.241pt}{0.400pt}}
\put(893.0,781.0){\rule[-0.200pt]{0.723pt}{0.400pt}}
\put(904,782.67){\rule{0.482pt}{0.400pt}}
\multiput(904.00,782.17)(1.000,1.000){2}{\rule{0.241pt}{0.400pt}}
\put(906,783.67){\rule{0.723pt}{0.400pt}}
\multiput(906.00,783.17)(1.500,1.000){2}{\rule{0.361pt}{0.400pt}}
\put(901.0,783.0){\rule[-0.200pt]{0.723pt}{0.400pt}}
\put(911,784.67){\rule{0.723pt}{0.400pt}}
\multiput(911.00,784.17)(1.500,1.000){2}{\rule{0.361pt}{0.400pt}}
\put(914,785.67){\rule{0.723pt}{0.400pt}}
\multiput(914.00,785.17)(1.500,1.000){2}{\rule{0.361pt}{0.400pt}}
\put(909.0,785.0){\rule[-0.200pt]{0.482pt}{0.400pt}}
\put(919,786.67){\rule{0.723pt}{0.400pt}}
\multiput(919.00,786.17)(1.500,1.000){2}{\rule{0.361pt}{0.400pt}}
\put(922,787.67){\rule{0.482pt}{0.400pt}}
\multiput(922.00,787.17)(1.000,1.000){2}{\rule{0.241pt}{0.400pt}}
\put(917.0,787.0){\rule[-0.200pt]{0.482pt}{0.400pt}}
\put(927,788.67){\rule{0.482pt}{0.400pt}}
\multiput(927.00,788.17)(1.000,1.000){2}{\rule{0.241pt}{0.400pt}}
\put(924.0,789.0){\rule[-0.200pt]{0.723pt}{0.400pt}}
\put(932,789.67){\rule{0.482pt}{0.400pt}}
\multiput(932.00,789.17)(1.000,1.000){2}{\rule{0.241pt}{0.400pt}}
\put(934,790.67){\rule{0.723pt}{0.400pt}}
\multiput(934.00,790.17)(1.500,1.000){2}{\rule{0.361pt}{0.400pt}}
\put(929.0,790.0){\rule[-0.200pt]{0.723pt}{0.400pt}}
\put(940,791.67){\rule{0.482pt}{0.400pt}}
\multiput(940.00,791.17)(1.000,1.000){2}{\rule{0.241pt}{0.400pt}}
\put(942,792.67){\rule{0.723pt}{0.400pt}}
\multiput(942.00,792.17)(1.500,1.000){2}{\rule{0.361pt}{0.400pt}}
\put(937.0,792.0){\rule[-0.200pt]{0.723pt}{0.400pt}}
\put(947,793.67){\rule{0.723pt}{0.400pt}}
\multiput(947.00,793.17)(1.500,1.000){2}{\rule{0.361pt}{0.400pt}}
\put(945.0,794.0){\rule[-0.200pt]{0.482pt}{0.400pt}}
\put(952,794.67){\rule{0.723pt}{0.400pt}}
\multiput(952.00,794.17)(1.500,1.000){2}{\rule{0.361pt}{0.400pt}}
\put(955,795.67){\rule{0.723pt}{0.400pt}}
\multiput(955.00,795.17)(1.500,1.000){2}{\rule{0.361pt}{0.400pt}}
\put(950.0,795.0){\rule[-0.200pt]{0.482pt}{0.400pt}}
\put(960,796.67){\rule{0.723pt}{0.400pt}}
\multiput(960.00,796.17)(1.500,1.000){2}{\rule{0.361pt}{0.400pt}}
\put(958.0,797.0){\rule[-0.200pt]{0.482pt}{0.400pt}}
\put(965,797.67){\rule{0.723pt}{0.400pt}}
\multiput(965.00,797.17)(1.500,1.000){2}{\rule{0.361pt}{0.400pt}}
\put(968,798.67){\rule{0.482pt}{0.400pt}}
\multiput(968.00,798.17)(1.000,1.000){2}{\rule{0.241pt}{0.400pt}}
\put(963.0,798.0){\rule[-0.200pt]{0.482pt}{0.400pt}}
\put(973,799.67){\rule{0.482pt}{0.400pt}}
\multiput(973.00,799.17)(1.000,1.000){2}{\rule{0.241pt}{0.400pt}}
\put(970.0,800.0){\rule[-0.200pt]{0.723pt}{0.400pt}}
\put(978,800.67){\rule{0.723pt}{0.400pt}}
\multiput(978.00,800.17)(1.500,1.000){2}{\rule{0.361pt}{0.400pt}}
\put(975.0,801.0){\rule[-0.200pt]{0.723pt}{0.400pt}}
\put(983,801.67){\rule{0.723pt}{0.400pt}}
\multiput(983.00,801.17)(1.500,1.000){2}{\rule{0.361pt}{0.400pt}}
\put(981.0,802.0){\rule[-0.200pt]{0.482pt}{0.400pt}}
\put(988,802.67){\rule{0.723pt}{0.400pt}}
\multiput(988.00,802.17)(1.500,1.000){2}{\rule{0.361pt}{0.400pt}}
\put(991,803.67){\rule{0.482pt}{0.400pt}}
\multiput(991.00,803.17)(1.000,1.000){2}{\rule{0.241pt}{0.400pt}}
\put(986.0,803.0){\rule[-0.200pt]{0.482pt}{0.400pt}}
\put(996,804.67){\rule{0.482pt}{0.400pt}}
\multiput(996.00,804.17)(1.000,1.000){2}{\rule{0.241pt}{0.400pt}}
\put(993.0,805.0){\rule[-0.200pt]{0.723pt}{0.400pt}}
\put(1001,805.67){\rule{0.723pt}{0.400pt}}
\multiput(1001.00,805.17)(1.500,1.000){2}{\rule{0.361pt}{0.400pt}}
\put(998.0,806.0){\rule[-0.200pt]{0.723pt}{0.400pt}}
\put(1006,806.67){\rule{0.723pt}{0.400pt}}
\multiput(1006.00,806.17)(1.500,1.000){2}{\rule{0.361pt}{0.400pt}}
\put(1004.0,807.0){\rule[-0.200pt]{0.482pt}{0.400pt}}
\put(1011,807.67){\rule{0.723pt}{0.400pt}}
\multiput(1011.00,807.17)(1.500,1.000){2}{\rule{0.361pt}{0.400pt}}
\put(1009.0,808.0){\rule[-0.200pt]{0.482pt}{0.400pt}}
\put(1016,808.67){\rule{0.723pt}{0.400pt}}
\multiput(1016.00,808.17)(1.500,1.000){2}{\rule{0.361pt}{0.400pt}}
\put(1014.0,809.0){\rule[-0.200pt]{0.482pt}{0.400pt}}
\put(1022,809.67){\rule{0.482pt}{0.400pt}}
\multiput(1022.00,809.17)(1.000,1.000){2}{\rule{0.241pt}{0.400pt}}
\put(1019.0,810.0){\rule[-0.200pt]{0.723pt}{0.400pt}}
\put(1027,810.67){\rule{0.482pt}{0.400pt}}
\multiput(1027.00,810.17)(1.000,1.000){2}{\rule{0.241pt}{0.400pt}}
\put(1024.0,811.0){\rule[-0.200pt]{0.723pt}{0.400pt}}
\put(1032,811.67){\rule{0.482pt}{0.400pt}}
\multiput(1032.00,811.17)(1.000,1.000){2}{\rule{0.241pt}{0.400pt}}
\put(1029.0,812.0){\rule[-0.200pt]{0.723pt}{0.400pt}}
\put(1037,812.67){\rule{0.482pt}{0.400pt}}
\multiput(1037.00,812.17)(1.000,1.000){2}{\rule{0.241pt}{0.400pt}}
\put(1034.0,813.0){\rule[-0.200pt]{0.723pt}{0.400pt}}
\put(1042,813.67){\rule{0.723pt}{0.400pt}}
\multiput(1042.00,813.17)(1.500,1.000){2}{\rule{0.361pt}{0.400pt}}
\put(1039.0,814.0){\rule[-0.200pt]{0.723pt}{0.400pt}}
\put(1047,814.67){\rule{0.723pt}{0.400pt}}
\multiput(1047.00,814.17)(1.500,1.000){2}{\rule{0.361pt}{0.400pt}}
\put(1045.0,815.0){\rule[-0.200pt]{0.482pt}{0.400pt}}
\put(1052,815.67){\rule{0.723pt}{0.400pt}}
\multiput(1052.00,815.17)(1.500,1.000){2}{\rule{0.361pt}{0.400pt}}
\put(1050.0,816.0){\rule[-0.200pt]{0.482pt}{0.400pt}}
\put(1057,816.67){\rule{0.723pt}{0.400pt}}
\multiput(1057.00,816.17)(1.500,1.000){2}{\rule{0.361pt}{0.400pt}}
\put(1055.0,817.0){\rule[-0.200pt]{0.482pt}{0.400pt}}
\put(1063,817.67){\rule{0.482pt}{0.400pt}}
\multiput(1063.00,817.17)(1.000,1.000){2}{\rule{0.241pt}{0.400pt}}
\put(1060.0,818.0){\rule[-0.200pt]{0.723pt}{0.400pt}}
\put(1068,818.67){\rule{0.482pt}{0.400pt}}
\multiput(1068.00,818.17)(1.000,1.000){2}{\rule{0.241pt}{0.400pt}}
\put(1065.0,819.0){\rule[-0.200pt]{0.723pt}{0.400pt}}
\put(1075,819.67){\rule{0.723pt}{0.400pt}}
\multiput(1075.00,819.17)(1.500,1.000){2}{\rule{0.361pt}{0.400pt}}
\put(1070.0,820.0){\rule[-0.200pt]{1.204pt}{0.400pt}}
\put(1080,820.67){\rule{0.723pt}{0.400pt}}
\multiput(1080.00,820.17)(1.500,1.000){2}{\rule{0.361pt}{0.400pt}}
\put(1078.0,821.0){\rule[-0.200pt]{0.482pt}{0.400pt}}
\put(1086,821.67){\rule{0.482pt}{0.400pt}}
\multiput(1086.00,821.17)(1.000,1.000){2}{\rule{0.241pt}{0.400pt}}
\put(1083.0,822.0){\rule[-0.200pt]{0.723pt}{0.400pt}}
\put(1091,822.67){\rule{0.482pt}{0.400pt}}
\multiput(1091.00,822.17)(1.000,1.000){2}{\rule{0.241pt}{0.400pt}}
\put(1088.0,823.0){\rule[-0.200pt]{0.723pt}{0.400pt}}
\put(1098,823.67){\rule{0.723pt}{0.400pt}}
\multiput(1098.00,823.17)(1.500,1.000){2}{\rule{0.361pt}{0.400pt}}
\put(1093.0,824.0){\rule[-0.200pt]{1.204pt}{0.400pt}}
\put(1103,824.67){\rule{0.723pt}{0.400pt}}
\multiput(1103.00,824.17)(1.500,1.000){2}{\rule{0.361pt}{0.400pt}}
\put(1101.0,825.0){\rule[-0.200pt]{0.482pt}{0.400pt}}
\put(1111,825.67){\rule{0.723pt}{0.400pt}}
\multiput(1111.00,825.17)(1.500,1.000){2}{\rule{0.361pt}{0.400pt}}
\put(1106.0,826.0){\rule[-0.200pt]{1.204pt}{0.400pt}}
\put(1116,826.67){\rule{0.723pt}{0.400pt}}
\multiput(1116.00,826.17)(1.500,1.000){2}{\rule{0.361pt}{0.400pt}}
\put(1114.0,827.0){\rule[-0.200pt]{0.482pt}{0.400pt}}
\put(1124,827.67){\rule{0.723pt}{0.400pt}}
\multiput(1124.00,827.17)(1.500,1.000){2}{\rule{0.361pt}{0.400pt}}
\put(1119.0,828.0){\rule[-0.200pt]{1.204pt}{0.400pt}}
\put(1129,828.67){\rule{0.723pt}{0.400pt}}
\multiput(1129.00,828.17)(1.500,1.000){2}{\rule{0.361pt}{0.400pt}}
\put(1127.0,829.0){\rule[-0.200pt]{0.482pt}{0.400pt}}
\put(1137,829.67){\rule{0.482pt}{0.400pt}}
\multiput(1137.00,829.17)(1.000,1.000){2}{\rule{0.241pt}{0.400pt}}
\put(1132.0,830.0){\rule[-0.200pt]{1.204pt}{0.400pt}}
\put(1142,830.67){\rule{0.482pt}{0.400pt}}
\multiput(1142.00,830.17)(1.000,1.000){2}{\rule{0.241pt}{0.400pt}}
\put(1139.0,831.0){\rule[-0.200pt]{0.723pt}{0.400pt}}
\put(1150,831.67){\rule{0.482pt}{0.400pt}}
\multiput(1150.00,831.17)(1.000,1.000){2}{\rule{0.241pt}{0.400pt}}
\put(1144.0,832.0){\rule[-0.200pt]{1.445pt}{0.400pt}}
\put(1157,832.67){\rule{0.723pt}{0.400pt}}
\multiput(1157.00,832.17)(1.500,1.000){2}{\rule{0.361pt}{0.400pt}}
\put(1152.0,833.0){\rule[-0.200pt]{1.204pt}{0.400pt}}
\put(1162,833.67){\rule{0.723pt}{0.400pt}}
\multiput(1162.00,833.17)(1.500,1.000){2}{\rule{0.361pt}{0.400pt}}
\put(1160.0,834.0){\rule[-0.200pt]{0.482pt}{0.400pt}}
\put(1170,834.67){\rule{0.723pt}{0.400pt}}
\multiput(1170.00,834.17)(1.500,1.000){2}{\rule{0.361pt}{0.400pt}}
\put(1165.0,835.0){\rule[-0.200pt]{1.204pt}{0.400pt}}
\put(1178,835.67){\rule{0.482pt}{0.400pt}}
\multiput(1178.00,835.17)(1.000,1.000){2}{\rule{0.241pt}{0.400pt}}
\put(1173.0,836.0){\rule[-0.200pt]{1.204pt}{0.400pt}}
\put(1185,836.67){\rule{0.723pt}{0.400pt}}
\multiput(1185.00,836.17)(1.500,1.000){2}{\rule{0.361pt}{0.400pt}}
\put(1180.0,837.0){\rule[-0.200pt]{1.204pt}{0.400pt}}
\put(1193,837.67){\rule{0.723pt}{0.400pt}}
\multiput(1193.00,837.17)(1.500,1.000){2}{\rule{0.361pt}{0.400pt}}
\put(1188.0,838.0){\rule[-0.200pt]{1.204pt}{0.400pt}}
\put(1201,838.67){\rule{0.482pt}{0.400pt}}
\multiput(1201.00,838.17)(1.000,1.000){2}{\rule{0.241pt}{0.400pt}}
\put(1196.0,839.0){\rule[-0.200pt]{1.204pt}{0.400pt}}
\put(1208,839.67){\rule{0.723pt}{0.400pt}}
\multiput(1208.00,839.17)(1.500,1.000){2}{\rule{0.361pt}{0.400pt}}
\put(1203.0,840.0){\rule[-0.200pt]{1.204pt}{0.400pt}}
\put(1219,840.67){\rule{0.482pt}{0.400pt}}
\multiput(1219.00,840.17)(1.000,1.000){2}{\rule{0.241pt}{0.400pt}}
\put(1211.0,841.0){\rule[-0.200pt]{1.927pt}{0.400pt}}
\put(1226,841.67){\rule{0.723pt}{0.400pt}}
\multiput(1226.00,841.17)(1.500,1.000){2}{\rule{0.361pt}{0.400pt}}
\put(1221.0,842.0){\rule[-0.200pt]{1.204pt}{0.400pt}}
\put(1234,842.67){\rule{0.723pt}{0.400pt}}
\multiput(1234.00,842.17)(1.500,1.000){2}{\rule{0.361pt}{0.400pt}}
\put(1229.0,843.0){\rule[-0.200pt]{1.204pt}{0.400pt}}
\put(1244,843.67){\rule{0.723pt}{0.400pt}}
\multiput(1244.00,843.17)(1.500,1.000){2}{\rule{0.361pt}{0.400pt}}
\put(1237.0,844.0){\rule[-0.200pt]{1.686pt}{0.400pt}}
\put(1252,844.67){\rule{0.723pt}{0.400pt}}
\multiput(1252.00,844.17)(1.500,1.000){2}{\rule{0.361pt}{0.400pt}}
\put(1247.0,845.0){\rule[-0.200pt]{1.204pt}{0.400pt}}
\put(1262,845.67){\rule{0.723pt}{0.400pt}}
\multiput(1262.00,845.17)(1.500,1.000){2}{\rule{0.361pt}{0.400pt}}
\put(1255.0,846.0){\rule[-0.200pt]{1.686pt}{0.400pt}}
\put(1273,846.67){\rule{0.482pt}{0.400pt}}
\multiput(1273.00,846.17)(1.000,1.000){2}{\rule{0.241pt}{0.400pt}}
\put(1265.0,847.0){\rule[-0.200pt]{1.927pt}{0.400pt}}
\put(1283,847.67){\rule{0.482pt}{0.400pt}}
\multiput(1283.00,847.17)(1.000,1.000){2}{\rule{0.241pt}{0.400pt}}
\put(1275.0,848.0){\rule[-0.200pt]{1.927pt}{0.400pt}}
\put(1293,848.67){\rule{0.723pt}{0.400pt}}
\multiput(1293.00,848.17)(1.500,1.000){2}{\rule{0.361pt}{0.400pt}}
\put(1285.0,849.0){\rule[-0.200pt]{1.927pt}{0.400pt}}
\put(1303,849.67){\rule{0.723pt}{0.400pt}}
\multiput(1303.00,849.17)(1.500,1.000){2}{\rule{0.361pt}{0.400pt}}
\put(1296.0,850.0){\rule[-0.200pt]{1.686pt}{0.400pt}}
\put(1314,850.67){\rule{0.482pt}{0.400pt}}
\multiput(1314.00,850.17)(1.000,1.000){2}{\rule{0.241pt}{0.400pt}}
\put(1306.0,851.0){\rule[-0.200pt]{1.927pt}{0.400pt}}
\put(1326,851.67){\rule{0.723pt}{0.400pt}}
\multiput(1326.00,851.17)(1.500,1.000){2}{\rule{0.361pt}{0.400pt}}
\put(1316.0,852.0){\rule[-0.200pt]{2.409pt}{0.400pt}}
\put(1339,852.67){\rule{0.723pt}{0.400pt}}
\multiput(1339.00,852.17)(1.500,1.000){2}{\rule{0.361pt}{0.400pt}}
\put(1329.0,853.0){\rule[-0.200pt]{2.409pt}{0.400pt}}
\put(1352,853.67){\rule{0.482pt}{0.400pt}}
\multiput(1352.00,853.17)(1.000,1.000){2}{\rule{0.241pt}{0.400pt}}
\put(1342.0,854.0){\rule[-0.200pt]{2.409pt}{0.400pt}}
\put(1365,854.67){\rule{0.482pt}{0.400pt}}
\multiput(1365.00,854.17)(1.000,1.000){2}{\rule{0.241pt}{0.400pt}}
\put(1354.0,855.0){\rule[-0.200pt]{2.650pt}{0.400pt}}
\put(1380,855.67){\rule{0.723pt}{0.400pt}}
\multiput(1380.00,855.17)(1.500,1.000){2}{\rule{0.361pt}{0.400pt}}
\put(1367.0,856.0){\rule[-0.200pt]{3.132pt}{0.400pt}}
\put(1393,856.67){\rule{0.482pt}{0.400pt}}
\multiput(1393.00,856.17)(1.000,1.000){2}{\rule{0.241pt}{0.400pt}}
\put(1383.0,857.0){\rule[-0.200pt]{2.409pt}{0.400pt}}
\put(1411,857.67){\rule{0.482pt}{0.400pt}}
\multiput(1411.00,857.17)(1.000,1.000){2}{\rule{0.241pt}{0.400pt}}
\put(1395.0,858.0){\rule[-0.200pt]{3.854pt}{0.400pt}}
\put(1429,858.67){\rule{0.482pt}{0.400pt}}
\multiput(1429.00,858.17)(1.000,1.000){2}{\rule{0.241pt}{0.400pt}}
\put(1413.0,859.0){\rule[-0.200pt]{3.854pt}{0.400pt}}
\put(1431.0,860.0){\rule[-0.200pt]{1.927pt}{0.400pt}}
\end{picture}